\documentclass[proceedings, preprint]{rmaa}

\usepackage{paralist}

\usepackage{psfrag,color}

\SetYear{2024}
\SetConfTitle{Workshop on Robotic Autonomous Observatories (AstroRob 2023) }

\title{New  Robotic Telescope: The big eye to observe the transient Universe} 

\author{
C.~M. Guti\'errez,\altaffilmark{1,2}
J. Barrera,\altaffilmark{1,2}
J. Bento,\altaffilmark{3}
D. Copley,\altaffilmark{3}
C.~M. Copperwheat,\altaffilmark{3}
F.~J. De Cos,\altaffilmark{4,5}
M. Escriche,\altaffilmark{1,2}
J.~J. Fern\'andez-Valdivia,\altaffilmark{1,2} 
A. P. Garner,\altaffilmark{3}
J. Gracia,\altaffilmark{4,6}
D.~G. Heffernan-Clarke,\altaffilmark{3}
H.~E. Jermak,\altaffilmark{3}
J. Le\'on Gil,\altaffilmark{1,2}
A.~M. McGrath,\altaffilmark{3}
C. Miossec,\altaffilmark{3}
A. Oria,\altaffilmark{1,2}
A. Ranjbar,\altaffilmark{3}
R. Rebolo,\altaffilmark{1,2}
C. Rodr\'\i guez-Pereira,\altaffilmark{4}
F. S\'anchez-Lasheras,\altaffilmark{4,7}
R.~J. Smith,\altaffilmark{3}
I.~A. Steele,\altaffilmark{3}
M. Torres,\altaffilmark{1,2}
}

\altaffiltext{1}{Instituto de Astrof\'\i sica de Canarias, E-38205
Tenerife, Spain}
\altaffiltext{2}{University of La Laguna, E-38206 Tenerife, Spain}
\altaffiltext{3}{Astrophysics Research Institute, Liverpool John Moores University,
IC2, Liverpool Science Park, 146 Brownlow Hill, Liverpool, L3 5RF,UK}
\altaffiltext{4}{Instituto Universitario de Ciencias y Tecnologías Espaciales de Asturias (ICTEA), c/ Independencia 13, 33004 Oviedo, Spain}
\altaffiltext{5}{Department of Prospecting and Exploitation of Mines, University of Oviedo, 33004 Oviedo, Spain}
\altaffiltext{6}{Department of Construction and Manufacturing Engineering, Oviedo University, c/ Pedro Puig Adam, 33203 Gijón, Spain}
\altaffiltext{7}{Department of Mathematics, Oviedo University, Oviedo, c/ Leopoldo Calvo Sotelo 18, 33007 Oviedo, Spain}

\shortauthor{Guti\'errez et al.}
\shorttitle{The New Robotic Telescope}

\listofauthors{C. M. Guti\'errez}
\indexauthor{Henney, W. J.}
\indexauthor{Collaborator, A.}
\indexauthor{Author, L.}

\abstract{
NRT is an international project to build and operate the world's largest robotic telescope. The telescope will have a segmented primary mirror with an equivalent diameter of 4 m, a set of simultaneously mounted optical and near-infrared instruments, and a response time of less than 30 seconds.  The project builds on the experience gained with the successful twenty-year operation of the Liverpool telescope, and with the GTC optics and control system. All of the above together with the excellent conditions for astronomical observation of La Palma, represents a solid base and guarantees that NRT will be one of the leading facilities in the field of time domain astronomy. This contribution will analyze the current status of the project with special emphasis on the development of its optics, and the plans for its construction and operation.} 

\resumen{NRT es un proyecto internacional para construir y operar el mayor telescopio rob\'otico del mundo. El telescopio tendr\'a~un espejo primario segmentado con un di\'ametro equivalente de 4 m, un conjunto de instrumentos \'opticos y de infrarrojo cercano montados simult\'aneamente, y un tiempo de respuesta inferior a 30 segundos.  El proyecto se basa en la experiencia adquirida con el exitoso funcionamiento durante veinte a\~nos del telescopio de Liverpool, y con la \'optica y el sistema de control del telescopio GTC. Todo ello, unido a las excelentes condiciones para la observaci\'on astron\'omica de La Palma, representa una base s\'olida y garantiza que el NRT ser\'a una de las instalaciones punteras en el campo de la astronom\'\i a en el dominio temporal. Esta contribuci\'on analiza el estado actual del proyecto, con especial \'enfasis en el desarrollo de su \'optica, y los planes para su construcci\'on y operaci\'on.}

\addkeyword{robotic telescopes}
\addkeyword{time-domain astronomy}

\begin{document}
\maketitle

\section{Introduction}

Astronomical observations in recent years are showing a changing face of the Universe, where sudden transient phenomena have been discovered on timescales that can be studied with the telescopes and instrumentation available nowdays. This has been made possible by advances in instrumentation and the design of increasingly versatile telescopes both on the ground and in space. Many of these telescopes allow efficient mapping of large regions of the sky and the access to unexplored ranges of the electromagnetic spectrum. These observations show a dynamic Universe spanning a wide range of spatial and temporal scales, encompassing from the discovery of new objects in the solar system, to galactic phenomena, such as novae, supernovae or tidal disruption events, to those of cosmological interest such as gamma-ray bursts or gravitational waves. Those phenomena open the door to the observation of new physical phenomena and to study the properties of matter and energy in extreme conditions, offering the possibility of discovering and extending our knowledge of the laws of nature.

A large number of telescopes have played a leading role in this progress, from those that act as discoverers of events, to those that are able to precisely locate and associate them with a particular astronomical source, to those that subsequently classify and characterise such events. As the number of discoveries of these transient phenomena is increasing, it is desirable to have telescopes that are fully dedicated to their study. Rapid localisation and classification is
essential to exploit the full potential of these phenomena, so such facilities must be able to respond rapidly to a given alert and provide the data quickly and efficiently to the community. Automated operation of all these processes is desirable. That was the starting point almost twenty years ago for the Liverpool telescope (Steele 2004)  which represented the first step towards a generation of large aperture telescopes operating in robotic mode and whose impact over the years has been demonstrated through a large number of leading scientific publications.

The new era of transient discoveries to be led among others by the Rubin Observatory (https://rubinobservatory.org), which will be able to make deep mappings of a representative part of the sky with cadences of a few days, makes the need for facilities capable of classifying and characterising these phenomena urgent. A detailed study of each of these phenomena would require a titanic observational task and an unfeasible number of facilities in the world. Therefore, robotic telescopes with relatively large apertures are essential to perform a selection of the most interesting cases and an optimal observation of them, allowing a rapid classification of the phenomena and their subsequent study with large telescopes. That was the motivation of the project presented in this contribution. Building on the experience gained during almost two decades of successful operation of the Liverpool telescope, we
 proposed to take the first step towards a new generation of 4-metre class robotic telescopes, with the New Robotic Telescope (NRT) project as the first representative and guide of this new generation. The concept study and the main scientific drivers were presented in Copperwheat et al. (2015). Here, we present an update to those presented in past years (Guti\'errez et al. 2019, 2021),  highlighting the current status of the project, its plans until first light, and emphasising the optical, mechanical and control system solutions adopted. 

\section{The project}

The technological development of the project began in 2018 promoted by 
the University of Liverpool (LJMU), in collaboration with the Instituto de Astrof\'\i sica de Canarias (IAC) and soon joined by the University of Oviedo (UoO).
The basic requirements were fully robotic operation, aperture size, operation in the optical and infrared ranges, a response time to an alert of interest of less than 30 seconds, a Ritchey-Chretien optical configuration with a field of view of 15 arcmin, and six focal stations allowing the simultaneous mounting of several instruments. The image quality requirement implied a deterioration with respect to seeing of no more than 0.2 arcseconds at 80 \% fraction of enclosed energy  for long exposures. 

The project has gone through its different design phases culminating in its  Preliminar Design Review (PDR)  Design Review) at the end of 2021, and is currently in the phase of detailed design in order to carrying out a Critical Design Review (CDR) in 2025. At the same time, the corresponding studies and environmental impact assessments are being carried out and the corresponding administrative permits are being obtained. The agreements for the construction and operation of the telescope are also being formalised, setting out the rights and obligations of each of the parties. We estimate that the first light will take place in 2028.

The telescope will be installed in the Roque de los Muchachos Observatory (La Palma, Spain), complementing and adding value to the existing facilities at the observatory, which make it one of the most important sites in the world for astronomical observations. The Observatory combines this excellent quality for observation with a set of logistical advantages, including the existence of an international airport just over an hour's drive from the observatory, which can be accessed by several paved roads that are open almost 365 days a year, as well as the existence of a set of infrastructures, technical support and specialised personnel that facilitate the effective operation of the telescope. To all this must be added the European Union (EU) own legal framework, which guarantees  stability and legal certainty.

Within the observatory, a location previously occupied by a now decomissioned facility (the Carlsberg Meridian Telescope) has been chosen to minimise the environmental and landscape impact and to optimise resources in an efficient way.  Atmospheric data collected over the years near to that location show very good conditions in terms of seeing (median 0.7 arcseconds), useful observing hours, humidity, sky darkness, etc. Figure~1 shows a view of such location at the Roque de los Muchachos Observatory.

For the building and enclosure, a functional design concept (Gradisar et al. 2022) has been pursued to house the telescope, as well as facilities such as workshops and storage areas, in line with the minimal interaction required for such a telescope. Figure 2 shows an artistic simulation of what the enclosure will look like. A clam-shell solution has been chosen. This solution allows full all-sky observation above a height of 20 degrees, facilitates operation and maintenance, reduces dome seeing by allowing rapid heat dissipation before observations start, and is a solution that has proven successful over many years at the Liverpool telescope. On the downside, the permanent and complete exposure of the telescope to the atmosphere must be taken into account in the design and operation of the telescope.

\begin{figure}[!t]
  \includegraphics[width=\columnwidth]{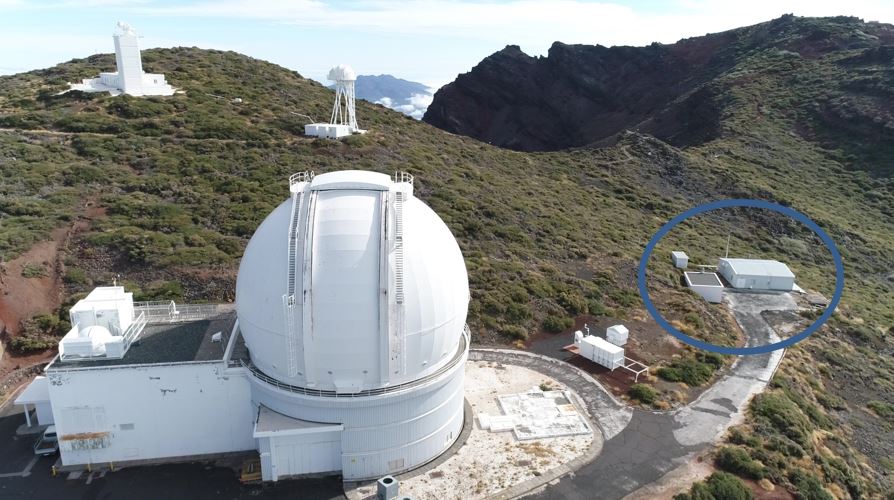}
  \caption{View of the Roque de los Muchachos Observatory  (La Palma, Spain) with the WHT telescope in front. The planned location of NRT is indicated by a blue ellipse.}
\end{figure}

\begin{figure*}[!t]
   \includegraphics[width=\columnwidth,height=4.5cm]{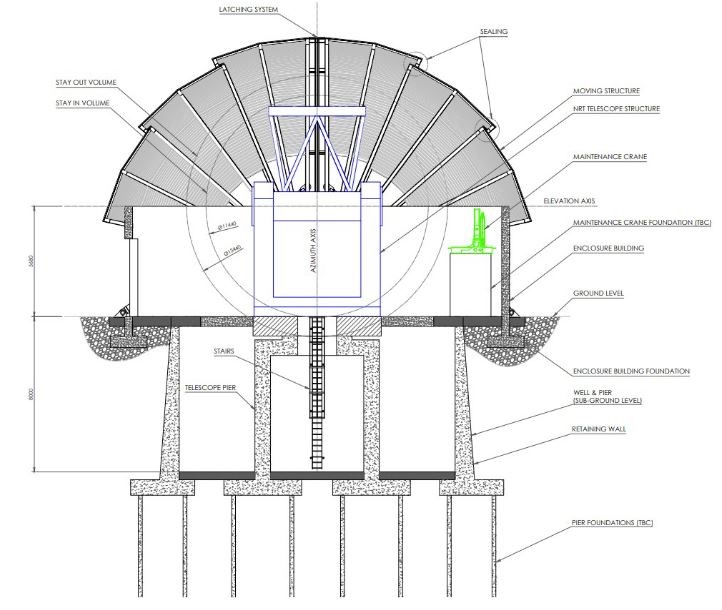}%
  \hspace*{\columnsep}%
  \includegraphics[width=\columnwidth,height=4.5cm]{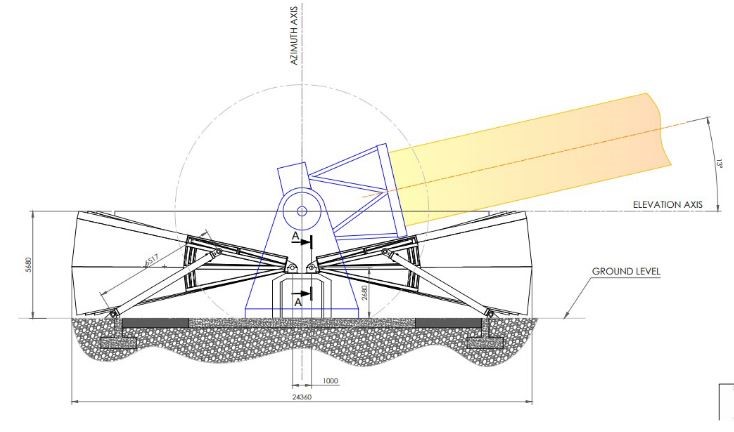}
  \caption{Enclosure of NRT.}
\end{figure*}

\section{Structure}

The design of the NRT structure has undergone multiple iterations and analyses to optimise its performance (Rodr{\'\i}guez Pereira et al. 2022). Special attention was paid to achieving a fast response time, sufficient stiffness, and quick settling time for the telescope in order to accurately observe and respond to quickly fading transient events. The original design based on a Serrurier truss showed the feasibility of the quick movements required for NRT, and determined that a locked rotor frequency resonance of 10 Hz was needed (Ranjbar et al. 2020). 

In subsequent analysis that design was changed to a Multibay Truss (Fig. 3), which provides greater stiffness at a lower weight compared to the original Serrurier Truss, and the primary cell was reconfigured. This change enabled a reduction in the overall weight of the optical support system by about 25 tonnes (33 \%~from the original design), thus lowering torque requirements for the drive systems while maintaining the same performance objectives of the Serrurier truss. In addition, this modification enabled a change in the configuration of the secondary structure, allowing a configuration of secondary support vanes overlapping the gaps in the primary inner segments, thus reducing the obstruction of the primary mirror. This Multibay truss provided also great accessibility for mirror segment handling, although it introduces some susceptibility to torsional motion, which was resolved by applying pre-stresses to the secondary mirror system vanes.

The drive system solution will use hydrostatic bearings and direct drive motors, which will be segmented in the azimuth axis. The low friction afforded by the hydrostatic bearings allows for good performance at low speeds for tracking, despite the high slewing speeds achieved during pointing operations. The direct drives will provide a smooth input torque, avoiding the noise a mechanical transmission would introduce into the system.

\begin{figure}[!b]
 \includegraphics[width=\columnwidth]{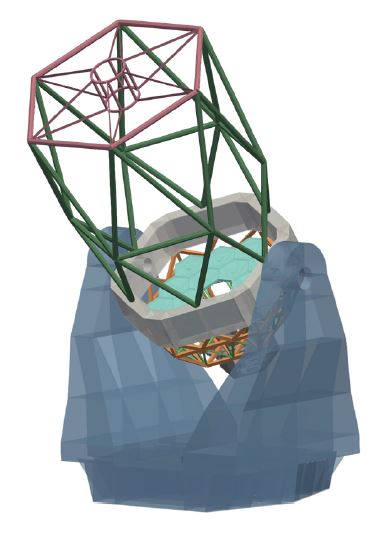}
  \caption{Artistic view of the concept for the mechanical structure of the NRT telescope.}
\end{figure}

To study the static behaviour a finite element model was applied obtaining a maximum offset (at the lowest elevation) between the primary and secondary mirrors of 0.083 mm and a defocus of 0.095 mm, which are within the requirements. In terms of dynamic behaviour, analyses have been carried out to estimate the natural frequencies of the structure which demonstrate its ability to move and stabilise within an acceptable range of vibration amplitude, in less than 30 seconds. As a whole, all these analyses show promising results in terms of low deformations and good optical, vibration, pointing, and tracking performance. In successive analyses we plan to include factors such as wind shaking or vibrations induced by the drive system during tracking.

\section{Optics and optomechanics}

The telescope will be a Ritchey-Chretien with F/10.6 in which the primary mirror consists of a set of 18 hexagonal mirrors of 96 cm placed in three concentric rings (Harvey et al. 2022). This gives a collecting surface which is equivalent to that of a 4 m diameter telescope. The secondary mirror is a lightened hyperbolic mirror with a diameter of 91 cm. Additionally, there is a mirror centred on the optical axis which allows the direct Cassegrain focal station to be exchanged for one of a set of 6 folded focal stations on which different instruments will be mounted simultaneously. 

Figure~3 shows the topological configuration adopted for the primary mirror of NRT.  Each of the segments will be an off-axis section of the nominal hyperboloid mounted on to an identical opto-mechanical support structure independently of their type. Each set of mirrors (of the same type) will be composed of 6 identical segments. This configuration requires three types of mirrors, due to the different radial positions of these groups with respect to the optical axis and the asphericity of the optical surface. 

\begin{figure}[!b]
  \includegraphics[width=\columnwidth]{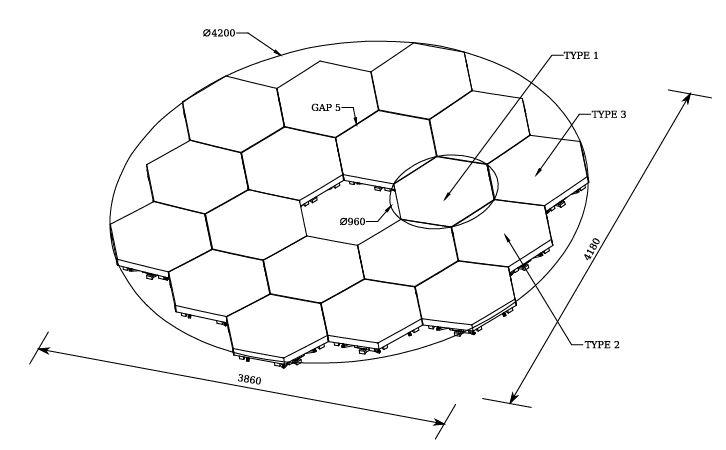}
  \caption{NRT primary mirror topology.}
\end{figure}

Although that type of segmented topology is typical of large telescopes such as GTC (www.gtc.iac.es), Keck (www.keckobservatory.org), and SALT (www.salt.ac.za), and of the new generation of large telescopes such as ELT (www.elt.eso.org) or TMT (www.tmt.org), there are some cases of smaller telescopes such as SEIMEI (Kurita et al. 2020) or LAMOST (www.lamost.org) that also use this type of solution. There were several reasons that led us to select a segmented configuration; firstly the segments allow for a weight reduction of several tons, in addition it allows the mirrors to be manufactured within the consortium; in fact, the IAC is developing an optics centre for the production of optical elements with  the equipment and metrology tools  to manufacture mirrors with diameters up to 1.5 m. The configuration chosen is similar to that of GTC, and it is therefore to be expected that the optomechanical and control system for NRT could be based on the solutions adopted by GTC. Furthermore, NRT will represent the pioneer for new generations of robotic telescopes with progressively larger apertures that will necessarily have to adopt segmented solutions above a given diameter. 

On the other hand, this solution poses additional problems such as the higher complexity of the positioning system, and the logistical difficulty of the recoating process in which a strategy of re-aluminisations by groups of three segments every several months is foreseen, instead of a single aluminisation every two years as would be necessary for a monolithic mirror. 

\begin{figure}[!t]
  \includegraphics[width=\columnwidth]{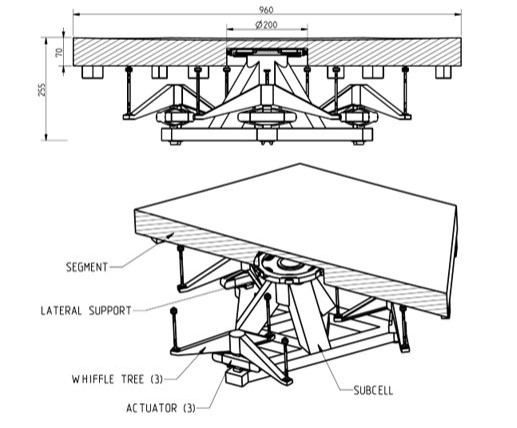}
  \caption{Primary  mirror segment support assembly.}
  \label{fig:simple3}
\end{figure}

Each primary mirror segment system will be composed of the substrate, and the axial and lateral pads as shown in Figure 5. The substrate will be a plano-concave hexagon with a mean size of 960 mm point-to-point and about 70 mm of maximum thickness (Oria 2022). The active optics implements actuators to move the mirrors (in piston, tip and tilt)  maintaining the shape of the M1 optical surface in the presence of different perturbations such as gravity, thermal changes or wind effects. The axial and lateral supports, and the actuators will be mounted on a common structure, the subcell that will hold both supports and actuators during segment handling. The axial load of each actuator will be distributed to three support points on the back plane of the substrate to reduce the segment gravity print-through. The load spreading element performing this function is called the whiffletree. The three different whiffletrees of the segment constitute its axial support system. 
The lateral support consists of a central diaphragm-like element located approximately at the segment center of gravity. The assembly is composed of a membrane, a monolithic outer frame including the radial vanes, a central hub and a mechanical safety limit to restrict the maximum motion range. This solution can take the lateral loads while remaining axially and tip/tilt compliant to decouple from the axial support.

An overview of the preliminary design of the secondary mirror system is shown in Figure 6. It is made up of the secondary mirror itself, its support system, and an active slow positioning stage to maintain the optical alignment with respect to the primary mirror. This unit can be precisely positioned in its six rigidbody degrees of freedom by means of a hexapod (see Fig. 6 bottom). As the secondary has to be supported by all the structural elements of the telescope, it is convenient to reduce its mass as much as possible to minimise the overall weight of the complete system. Figure 6 (top) shows the back side of the secondary mirror and the geometrical solution proposed for this weight reduction.

\begin{figure}[!t]
  \includegraphics[width=\columnwidth]{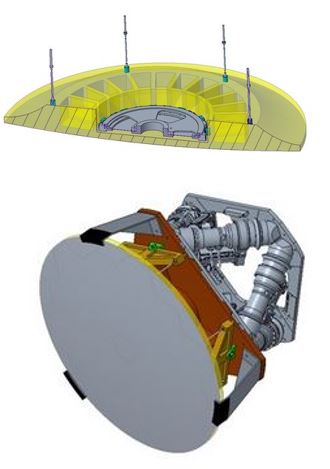}
  \caption{NRT secondary mirror system.}
\end{figure}

\section{Control system}

The NRT project implements a software architecture (Bento et al. 2022a) designed to support high-precision astronomical observations and user interactions with the telescope. 
A distributed architecture across two environments optimizes performance and accessibility, allowing for reliable nightly fully robotic operations and remote user interactions. A first environment, located on the observatory  forms the basis for a reliable software platform for nightly fully robotic operations. Containerized applications (Fern\'andez-Valdivia et al. 2022) are used to manage and process data directly from the telescope, enabling real-time analysis and immediate feedback on observations. The second environment will be hosted on Google Cloud, providing secure and scalable access to telescope data, analytical tools, and computational resources. 

GitLab\footnote[1]{https://about.gitlab.com/} is leveraged for version control, providing efficient code management and collaboration, while DevOps pipelines automate building, testing, and deployment processes. This automation is crucial for rapid iterate on the software, and  reduces the time and effort required to bring new features and updates to the infrastructure. Harbor\footnote[2]{https://goharbor.io/} is used as an artifact repository, ensuring secure storage and management of Docker images\footnote[3]{https://hub.docker.com/} and Helm charts\footnote[4]{https://helm.sh/} used in Kubernetes deployments. 

The software control topology is split into three layers: the Science Operations Data Centre (SODC), the Robotic Control System (RCS), and the Telescope Level Systems (TLS). The SODC layer (Jermak et al. 2022) handles user-facing functionality, including proposal submission, observation planning, data reduction pipelines, and access to data archives. 
The RCS layer is the equivalent to a human astronomer, making scientific decisions based on live conditions, prioritizing and scheduling observations, and providing fast-follow-up functionality. The Master Control Process acts as the ultimate decision maker within the RCS layer and communicates with the SODC layer through an application programming interface (API). 

The TLS layer is the link between the low-level hardware and the RCS. The TLS encompasses everything that directly controls hardware within the observatory and therefore, handles control of slewing, pointing, segmented mirror alignment, dome control and instrument interface. It contains the observing engine which allows targets requested by the RCS to be captured. The TLS is based (Bento et al. 2022b) on the existing GTC control system (GCS) that allows a ready-made functionality for key telescope behaviours such as segmented mirror alignment. 
Another key benefit of the GCS is that devices can be created to communicate to a wide variety of other systems and field busses. This means that both, instruments and control hardware can be operated directly by GCS without incorporating low level control within the TLS. 

Finally, the architecture for the low-level hardware control utilizes a distributed industrial network with Beckhoff PLCs and Ethercat data bus for robust and reliable control and communication between systems. The PLCs are controlled by a Kubernetes Cluster using the OPCUA\footnote[5]{https://opcfoundation.org/about/opc-technologies/opc-ua/} protocol.

\section{Conclusions}

NRT is an ambitious initiative that is set to play a key role in the new era of time domain astronomy, acting both as a pioneer and as an example of a generation of new telescopes extending robotic operation from the current 2-metre to 4-metre telescopes. This contribution has presented the motivation, objectives and main features of the project, as well as its current status  and the plans until its commissioning. The main features of the project are its relatively low cost of construction and operation, and its high efficiency and versatility that will allow it to address studies of a large number of transient phenomena by acting as a necessary link between the discovery instruments and the larger, traditionally operated telescopes that can carry out more detailed studies. With this approach, NRT aims to develop technological standards to guide this fascinating adventure to understand the world of transient phenomena in the universe.

\section*{Acknowledgements}

This Project has received funding from the Canary Islands Government under a direct grant awarded on grounds of public interest (Order Nº 185 August 2017). We also thank the support of the Spanish Ministerio de Ciencia, Innovaci\'on y Universidades, Fondo Europeo de Desarrollo Regional (FEDER), Agencia Canaria de Investigaci\'on, Innovaci\'on y Sociedad de la Informaci\'on (ACIISI) and the programmes Canarias Avanza con Europa and Tenerife Innova, Cabildo de Tenerife. The project is also supported by a UKRI grant from STFC (ST/V003828/1) and by the University of Oviedo, Project FUO-20-362.
\

\end{document}